\documentclass[aps,prl,
preprint
,showpacs
,groupedaddress
,superscriptaddress
,nobibnotes
]
{revtex4-1}
\usepackage{amsmath,amsthm,amssymb}
\usepackage{array}
\usepackage{subfigure}
\usepackage{graphicx}
\usepackage{fancyhdr}
\usepackage{color}
\usepackage{extarrows}
\usepackage{enumerate}
\usepackage{epstopdf}
\usepackage{bm}
\usepackage[colorlinks,
          linkcolor=black,
            citecolor=black,
            urlcolor=blue
           ]{hyperref}
\usepackage{natbib}
\usepackage{comment}

\usepackage{xr}
\begin{document}
\title{Resonant confinement of excitonic polariton and ultra-efficient light harvest in artificial photosynthesis}
\author{Yong-Cong Chen}
\email{chenyongcong@shu.edu.cn}
\affiliation{Shanghai Center for Quantitative Life Sciences \& Physics Department, Shanghai University, Shanghai 200444, China}
\author{Bo Song}
\affiliation{Shanghai Key Lab of Modern Optical System, School of
Optical-Electrical Computer Engineering, University of Shanghai for Science and Technology, Shanghai 200093, China}
\author{Anthony J. Leggett}
\affiliation{Shanghai Center for Complex Physics, Department of Physics and Astronomy, Shanghai Jiao Tong University, Shanghai 200240, China}
\affiliation{Department of Physics, University of Illinois at Urbana-Champaign, Urbana, IL 61801, USA}
\author{Ping Ao}
\affiliation{Shanghai Center for Quantitative Life Sciences \& Physics Department, Shanghai University, Shanghai 200444, China}
\author{Xiaomei Zhu}
\email[Corresponding author: ]{xiaomeizhu@shu.edu.cn}
\affiliation{Shanghai Center for Quantitative Life Sciences \& Physics Department, Shanghai University, Shanghai 200444, China}
\begin{abstract}
We show that in the recent artificial light-harvesting experiment 
[Angewandte Chemie Intl. Ed. 55, 2759 (2016)]
on organic nanocrystals self-assembled from difluoroboron chromophores, the spontaneous emission of an excited pigment should undergo a two-step process. It would first decay to an excitonic polariton confined by cavity resonance via strong photon-exciton coupling. The captive intermediate could then funnel the energy directly to doped acceptors, leading to the observed over 90\% transfer efficiency at less than 1/1000 acceptor-donor ratio. Theoretical, parameter-free analyses are in quantitative agreement with the experiment.
\end{abstract}
\pacs{81.16.Dn, 78.67.Bf, 42.50.Pq, 42.25.Gy}

\maketitle

\renewcommand{\vec}[1]{\mathbf{#1}}
\newcommand{\bmu}{\mbox{\boldmath $\mu$}}
\newcommand{\cmmnt}[1]{\ignorespaces}

In the photosynthesis of plants and bacteria, the unusual high efficiency (over 95\%) in the energy funnelled from a large number of antenna chromophores to a single collector in the reaction center for the antenna has always attracted a great deal of research interest\cite{doi:10.1021/acs.chemrev.6b00002}. Away from the living cells, efforts have been devoted to construct artificial light-harvesting systems that can emulate the success of their natural counterparts. Recently, one such success based on nanocrystals of difluoroboron chromophores was reported by Chen et~al.\cite{ANIE:ANIE201510503}, with the transfer efficiency reaching 95\%. We suggest in this letter that the high efficiency is a consequence of quantum confinement of excitonic polaritons via coherent exciton-photon coupling.

The nanocrystals in the experiment were assembled from a difluoroboron $\beta$-diketonate (BF$_{2}$dfbk) derivative BF$_{2}$bcz (C$_{31}$H$_{25}$BF$_{2}$N$_{2}$O$_{2}$ plus a solvent CHCl$_{3}$ for crystal assembly). It is chosen for its high fluorescence and makes up the antenna chromophores (donors). The collecting agents (acceptors) co-assembled into the crystal in small ratios are molecules of similar structure, BF$_{2}$cna or BF$_{2}$dan. The emission spectrum of BF$_{2}$bcz is in green light and has a large overlap with the absorption spectra of the acceptors. The latter in turn emit in red light, making the energy transfer vivid to the naked eye by color changes . The nanocrystals of BF$_{2}$bcz were reported to have smooth surfaces and uniform sizes of 400 - 600 nm in thickness and 5 - 7 $\mu$m in length\cite{ANIE:ANIE201510503,ADFM:ADFM201700332}.

All the molecules are typical D-A-D (electron donor-acceptor-donor) fluorophores, with the excited states showing strong ICT (intramolecular charge transfer) character. For BF$_{2}$bcz, the HOMO (highest occupied molecular orbital) state is concentrated on the two carbazole arms and the LUMO (lowest unoccupied molecular orbital) state is located on the difluoroboron moiety (BF$_{2}$)\cite{MARGAR2016241,Zhang2017143} connecting the arms. This results in a large dipole matrix element between the states.  The light harvesting process involves mainly  transitions between the HOMO and the LUMO electronic states. An excited donor molecule from UV illumination relaxes to the LUMO state after internal conversion. The energy was found to transfer to the LUMO state of a doped acceptor starting with a minute $10^{-6}$ acceptor-donor ratio and saturating at less than $10^{-3}$ doping with 95\% efficiency.

To explain the results by a diffusion-based mechanism would be difficult. In such a process, a localized exciton (excited electronic state on a molecule) would migrate to a neighboring molecule via e.g. FRET (F\"{o}rster resonance energy transfer)\cite{doi:10.1146/annurev-conmatphys-020911-125126}. The transfer rate would be limited by the nature of the random walks and the relatively short exciton lifetime ($\tau\sim3$ ns). The room-temperature hopping time is typically picoseconds or longer\cite{doi:10.1021/acs.chemrev.6b00002,doi:10.1080/15421407208083249,doi:10.1063/1.1676546}.  Nevertheless, in the experiment $\sim 10$\% energy transfer was observed at $\sim 10^{-6}$ acceptor-donor ratio. This would imply that an exciton needed to move $\sim 10^5$ steps with an average hopping time $\sim 10$ fs, two orders of magnitude faster than the typical scale. Hence the diffusion process could hardly account for the observed transfer efficiency. An alternative aggregate mechanism ought to exist in which the localized excitation should be converted into a form which propagates much faster than the F\"{o}rster mechanism and is sufficiently long-lived.

We can look into the opposite scenario in which a Mott-Wannier exciton could possibly propagate ballistically via the conduction and valence bands formed by virtue of the crystalline periodicity. Note that HOMO and LUMO derived bands in organic crystals usually have a small dispersion range, $\Delta E < 0.5$ eV (e.g. \cite{Koller351,doi:10.1063/1.3232205}).
For a Wannier exciton with wave amplitudes extended over many sites to gain the full energy advantage $\sim\Delta E$ from the bands, it would have to overcome the electron-hole Coulombic binding energy $\sim e^2/(4\pi\epsilon_0)(N/V)^{1/3} \sim 1$ eV on a localized exciton (cf. Table 1 in Ref.~\cite{NAYAK201342}). Hence the latter would be energetically favored. In the current context, crystalline fluorescence quantum yield remained high and the non-radiative decay rate stayed the same (cf. Part C of the supplementary infomation\cite{supplementary}). Large inter-molecular charge separations were evidently insignificant.

We can turn attention to the exciton bands where an electron-hole pair acts as a tightly bound entity\cite{doi:10.1146/annurev-physchem-040513-103654}. The overlapping of wave functions between neighboring sites is further reduced as it is now a product of the one between the HOMO states with that from the LUMO states\cite{doi:10.1021/acs.chemrev.6b00002}. The exciton coupling $J$ is typically $\sim 200 $ cm$^{-1}$ or $0.025$ eV\cite{doi:10.1021/acs.jpcc.6b01298,doi:10.1063/1.4955493}. On such an energy scale the thermal fluctuations at room temperature would destroy the ballistical propagation by excitons alone.

That leads us to the main proposal in this work: The sought aggregate mechanism may be provided by a resonance-induced exciton-polariton subject to self confinement. In an organic (molecular) crystal, the exciton states across the entire crystal can interact coherently with the photons to form compound excitations known as exciton-polaritons\cite{PhysRev.112.1555}.  Indeed light propagating as polaritons has been reported in many organic crystal waveguides\cite{PhysRevLett.105.067401,ADMA:ADMA201305913}.  For a crystal with dimensions comparable to the photon wavelength, polariton standing waves can become important. They can be trapped inside by total internal reflections at large $|\vec{k}|$ where the lower-branch polariton dispersion is capped by the bare exciton energy. The trapped states fall into an excitonic region in which the photon composition is small. If a majority of the wave modes at the emission spectrum are captive then a confined polariton becomes the intermediate state when a local exciton undergoes spontaneous emission. The polariton can in turn be absorbed efficiently by a doped acceptor.

The conjectured ``trapped exciton-polariton'' with the photon composition small, is an ideal candidate for the new form of energy speculated above. On one hand the strong exciton-photon coupling reaches the order $\sim1$ eV as a result of densely packed donor molecules with a large magnitude of the dipole matrix element. This allows the photon field to induce coherent response from the donor molecules across the entire crystal.  The photon-mediated mechanism can overcome the effects of static and/or dynamic disorder ($\sim 0.2$ eV) in localizing the exciton. The resulting hybrid modes offer the dominant channel to which a localized exciton can escape. Though each polariton has a relatively small photon composition, the reduction to the spontaneous emission rate is offset by an increase in the number of large $|\vec{k}|$ modes.  Together they turn the propagation of the excited energy into a ballistic form. On the other hand, the smallness of the photon composition suppresses the ``geometrical'' leaking rate out of the nanocrystal via reduced group velocities due to the saturated dispersion. The long confinement time ($\sim 5$ ps) is essential for a minute quantity of doped acceptors to compete against the leaking and achieve the high energy transfer rate.

We now proceed to justify the hypothesis. We will first show quantitatively how the resonant confinement arises inside the crystal based on a full quantum mechanical analysis. Consider the electro-magnetic waves inside a rectangular isotropic dielectric medium $L_1 \sim L_2 \sim \lambda_e$, $L_3 \sim 10\lambda_e$, with $\lambda_e\approx 500$ nm being the (vacuum) wavelength of the exciton emission. The dimensions are chosen to resemble the nanocrystals used in the experiment. Let the electric field
\begin{equation}\label{maxwell0}
 \vec{E}(\vec{r}, t) = \vec{U}_{\vec{k}\alpha}(\vec{r})\cos(\omega_{\vec{k}} t), \;\;\;  \epsilon_b\omega_{\vec{k}}^{2} = \vec{k}^{2}c^{2}
\end{equation}
be a standing wave inside the cavity with the background relative permittivity $\epsilon_b$ excluding the contribution from the excitons. We have
\begin{eqnarray}\label{maxwell}
\nonumber \vec{U}_{\vec{k}\alpha}(\vec{r}) &=& (u_{x\vec{k}\alpha}U_{x\vec{k}}, u_{y\vec{k}\alpha}U_{y\vec{k}}, u_{z\vec{k}\alpha}U_{z\vec{k}}) \\
\vec{u}_{\vec{k}\alpha}&=& (u_{x\vec{k}\alpha}, u_{y\vec{k}\alpha}, u_{z\vec{k}\alpha}) \;\;\; \alpha = 1, 2
\end{eqnarray}
\begin{align}\label{polarization}
\nonumber \vec{k} =(\pi l_1/L_1, \pi l_2/L_2, \pi l_3/L_3,) \;\;\; & l_i = 0, 1, 2, 3, \dots \\
          \tilde{\vec{k}} = \vec{k}/|\vec{k}| = \vec{u}_{\vec{k}1} \times \vec{u}_{\vec{k}2}\;\;\;
        & \vec{u}_{\vec{k}\alpha} \cdot \vec{u}_{\vec{k}\alpha'} = \delta_{\alpha\alpha'}
\end{align}
where $\vec{u}_{\vec{k}\alpha}$'s and $\vec{k}$ are respectively the polarization and wave vectors.
Explicit wave forms $U_{\gamma\vec{k}}(\vec{r})$ ($\gamma=x, y, z$) can be found in Ref.~\cite{Morse_Feshbach} and the supplementary information\cite{supplementary} for the present work.

The standing waves (\ref{maxwell}) are in essence linear superpositions of 8 plane waves with mirror reflections of $\vec{k}$ in some particular way. However, as the photon-exciton coupling turns on the energy reduction in the lower branch of the hybrid entity can bound the EM field outside to the cavity surfaces, forming evanescent waves. A wave is trapped inside when, with $\vec{k}_{\|}$ the tangent component of the wave vector, the inequality
\begin{equation}\label{criteria}
\vec{k}_{\|}^2 c^2 > \epsilon_s (E_{\vec{k}}/\hbar)^2 
\end{equation}
holds on all the surfaces. Here $\epsilon_s$ is the dielectric constant of the surrounding medium and $E_{\vec{k}}$ is the exciton-polariton energy. When a hybrid mode satisfies (\ref{criteria}) it becomes isolated from and only weakly interacts with the surroundings. The confined state can be properly studied by including the evanescent waves as a whole,  with a zero field boundary condition at infinity. As discussed in the supplementary information\cite{supplementary}, the wave forms \cmmnt{(\ref{S_maxwell1})}(S4) can reasonably represent the eigen state of the electric field inside. In addition, we will neglect the surface component when the bulk of the exciton-photon coupling is considered. This implicitly sets the lower bound on the crystal dimensions for the analyses below to remain valid.

To obtain $E_{\vec{k}}$ we need a full quantum mechanical modeling of the electro-magnetic interaction with the excitons as proposed by Hopfield\cite{PhysRev.112.1555}. To start with, one can quantize the macroscopic Maxwell equations using these eigenmodes. We shall not go through the full length of justification. For a given $\vec{k}$, the photon Hamiltonian and the electric field operator inside the cavity are given by (see e.g. \cite{PhysRevA.50.1830}),
\begin{eqnarray}\label{quantization}
  \hat{H}_{p\vec{k}} & = & \sum_{\alpha}\hbar\omega_{\vec{k}}(\hat{a}_{\vec{k}\alpha}^{\dag}\hat{a}_{\vec{k}\alpha} + \frac{1}{2})\\
  \hat{\vec{\mathcal{E}}}_{\vec{k}}(\vec{r}) & = & \sum_{\alpha}f_{\vec{k}}(\hat{a}_{\vec{k}\alpha} + \hat{a}_{\vec{k}\alpha}^{\dag})\vec{U}_{\vec{k}\alpha}(\vec{r})
\end{eqnarray}
where $f_{\vec{k}} = \sqrt{\hbar\omega_{\vec{k}}/(2\epsilon_0\epsilon_b)}$ and the phase of $\hat{a}_{\vec{k}\alpha}$ is chosen for symmetric $\hat{\vec{\mathcal{E}}}$.

The nanocrystals were illuminated by continuous UV light or weak laser pulses in time-resolved single photon counting measurements. The number of excitons in a crystal at a given time is estimated to be very small ($\sim 10$). Hence it is sufficient to consider the single exciton limit. We can approximate an excited molecule by a harmonic oscillator $\hat{b},\hat{b}^{\dag}$ restricted to its zeroth and first level, $E_1 = \hbar\omega_e \approx 2.5$ eV (for $\lambda_e \approx 500$ nm), namely the one excitation manifold. Let the position of the $j$th molecule be $\vec{R}_j$ and the dipole matrix element between the ground and the excited states be $\vec{d}=\langle g|\hat{\vec{r}}|e\rangle$, then under a rotating wave approximation (e.g. \cite{HEPP1973360,PhysRevLett.93.226403}) we have the exciton-photon interaction Hamiltonian
\begin{eqnarray}\label{interaction}
  \hat{H}_{I\vec{k}} & = & \sum_{j, \alpha}ef_{\vec{k}}\left[
  \vec{U}_{\vec{k}\alpha}(\vec{R}_j)\cdot(\vec{d}^{\ast}\hat{b}_{j}^{\dag}\hat{a}_{\vec{k}\alpha} + \vec{d}\hat{a}_{\vec{k}\alpha}^{\dag}\hat{b}_{j})
  \right] \\
\nonumber \empty & \empty & [\hat{b}_i,\hat{b}_j^{\dag}] = \delta_{ij}, \;\;\; i,j = 1, 2, \cdots, N
\end{eqnarray}
By taking all molecules with the same matrix element, we have ignored the minor orientation difference between the two groups of molecules in the crystal.

Similar to Refs.~\cite{PhysRev.93.99,PhysRev.112.1555}, we can construct a standing-wave (SW) superposition of the identical electronic states from all the molecules,
\begin{align}\label{coherent}
  \hat{b}_{z\vec{k}}^{\dag} &  = \sqrt\frac{V}{N}\sum_j\left[U_{z\vec{k}}(\vec{R}_j)\hat{b}_{j}^{\dag}\right] \\
  \empty & |\phi_{z\vec{k}}\rangle  = \hat{b}_{z\vec{k}}^{\dag}|0\rangle \;\;\; [\hat{b}_{z\vec{k}},\hat{b}_{z\vec{k}}^{\dag}] = 1
\end{align}
Two more SW states $|\phi_{x\vec{k}}\rangle$ and $|\phi_{y\vec{k}}\rangle$ can be likewise defined respectively for the $x$ and $y$ components. The three SW states are mutually orthogonal (for large $N$), though SW states from different $\vec{k}$'s can overlap due to over-completeness of the constructions.

In terms of the coherent SW states\cite{PhysRev.93.99, HEPP1973360}, the non-interactive Hamiltonian for a given $\vec{k}$ is
\begin{equation}\label{hamiltonian}
 \hat{H}_{0\vec{k}} = \sum_{\gamma}\hbar\omega_e\hat{b}_{\gamma\vec{k}}^{\dag}\hat{b}_{\gamma\vec{k}} +
 \sum_{\alpha}\hbar\omega_{\vec{k}}\hat{a}_{\vec{k}\alpha}^{\dag}\hat{a}_{\vec{k}\alpha}
\end{equation}
and the interaction term (\ref{interaction}) becomes
\begin{equation}\label{interaction1}
 \hat{H}_{I\vec{k}} = \sum_{\alpha, \gamma}\left[g_{\vec{k}}
            \tilde{\vec{d}}_{\gamma}^{\ast} \cdot \vec{u}_{\vec{k}\alpha}\hat{b}_{\gamma\vec{k}}^{\dag}\hat{a}_{\vec{k}\alpha} + h.c.\right] \\
\end{equation}
\begin{equation}\label{coupling}
 g_{\vec{k}} = e|\vec{d}|f_{\vec{k}}\sqrt{N/V}, \;\;\; \tilde{\vec{d}}_{\gamma} = \vec{d}_{\gamma}/|\vec{d}|
\end{equation}

The scale of the exciton-photon interaction is crucial. From Table 1 of Ref.~\cite{ADFM:ADFM201700332}, a monomer in DCM solution (CH$_2$Cl$_2$, refractive index 1.42) has $\tau_f = 2.12$ ns, $\Phi_f=0.80$ on the fluorescence lifetime and quantum yield, which translates to $|\vec{d}|\sim 2.0$ \AA\; via the standard spontaneous emission rate formula. The crystal has 4 molecules in a unit cell of size $2823$ \AA$^3$ (Table S2 of Support Information in \cite{ANIE:ANIE201510503}), hence $N/V=1.4\times 10^{-3}/$ \AA$^{3}$. At the resonance energy $\hbar\omega_{\vec{k}}=\hbar\omega_e$, $\epsilon_b\sim 2$, the coupling constant $g_{\vec{k}}$ evaluates to $\sim0.8$ eV. Note that by taking many parameters from the monomers in a solution, we implicitly assume that the crystalline electronic properties do not undergo significant changes upon self-assembly from the solution.

The photon-exciton coupling substantially exceeds the inter-molecular dipole-dipole coupling. The contribution from the latter to the exciton dispersion was approximately calculated in Eq.~(23) of Ref.~\cite{PhysRev.112.1555} (it appears to have a sign error and is valid up to a factor that depends on the lattice symmetry).  The matrix element is diagonal but orientation dependent in the plane wave representation,
\begin{equation}\label{dipole-dipole}
  D_{\vec{k}} = \frac{Ne^2|\vec{d}|^2}{3\epsilon_0\epsilon_b V}\left[1 - 3|\tilde{\vec{k}}\cdot\tilde{\vec{d}}|^2\right]
\end{equation}
The coefficient evaluates to $0.35/\epsilon_b \approx 0.17$ eV. Moreover, a standing wave is a composition of 8 plane waves over mirror reflections, the leading correction to $E_{\vec{k}}$ should be further reduced after average over the directions.

The two-state truncation employed above requires further scrutiny in the presence of local phonons. The latter are the low-lying vibrational modes associated with the electronic excitation. Their energy scale can be estimated from the Stokes shift $2\hbar\omega_s\approx 0.24$ eV in the fluorescence spectra (from Figure S8 of Ref.~\cite{ANIE:ANIE201510503}). In terms of the vertical transition picture the ground state can be excited into a number of vibronic (an electronic + a vibrational) states. Schematically, the Franck-Condon factor on the dipole element $\vec{d}$  for the transition to the $\nu$-th vibrational mode $|\nu\rangle$ has the form
\begin{equation}\label{franck-condon}
S_{\nu} = \exp(-g^2/2)\frac{g^{\nu}}{\sqrt{\nu!}} < 1
\end{equation}
where $g$ is some electron-phonon coupling constant.  We can introduce an effective excited state as a weighted superposition of all the vibrational states
\begin{equation}\label{effective-excited}
|e\rangle = \sum_{\nu} S_{\nu} |\nu\rangle
\end{equation}
which restores the transition dipole element back to its full oscillation strength. Furthermore, any other state orthogonal to (\ref{effective-excited}) will have a vanishing dipole matrix element, hence will not be coupled to the photons. This justifies the two-state scheme provided that the energy spread of the vibronic states $\hbar\omega_s\ll g_{\vec{k}}$ and the effective excited state energy is elevated to $\hbar\omega_e+\hbar\omega_s$.

Now the Hamiltonian $(\hat{H}_{0\vec{k}}+\hat{H}_{I\vec{k}})$ from Eqs~(\ref{hamiltonian}) and (\ref{interaction1}) can be readily diagonalized. The details are presented in the supplementary information\cite{supplementary}, Eqs.~\cmmnt{(\ref{S_dimensionless})}(S15) to \cmmnt{(\ref{S_groupv})}(S20). Under the criteria \cmmnt{(\ref{S_criteria1})}(S6), we are interested in where most captive modes reside and the corresponding energy drop $\Delta E_{\vec{k}} = \hbar\omega_e - E_{\vec{k}}$ from Eq.~\cmmnt{(\ref{S_isotropic})}(S19),  the photon composition factor $A_{\text{ph}}$ from Eq.~\cmmnt{(\ref{S_particle})}(S18), and the group velocity $v_{g}$ from Eq.~\cmmnt{(\ref{S_groupv})}(S20).
For the specific dimensions considered here $k_x$, $k_y$ can only take discrete values $\sim n k_e/2$, $k_e = \omega_e/c = (2\pi)/{\lambda_e}$ whereas $k_z$ is quasi-continuous. For $\epsilon_s$, $\epsilon_b \approx 2$ and if we set $k_x = 3k_e/2$, $k_y = 3k_e/2$ then $k_z \geq 0$ is free to take any values. For $k_x = k_e$, $k_y = 3k_e/2$ or $k_x = 3k_e/2$, $k_y = k_e$, we need $k_z \geq 0.9k_e$. This gives $\Delta E_{\vec{k}} \approx 0.22$ eV, $\Delta E_{\vec{k}}/g_{\vec{k}} \approx 0.24$, $A_{\text{ph}} \approx 0.15$, $v_{g} \approx 0.20 c/\sqrt{\epsilon_b}$. For $k_x = k_e$, $k_y = k_e$, $k_z \geq 0.7k_e$, we have $\Delta E_{\vec{k}} \approx 0.35$ eV, $\Delta E_{\vec{k}}/g_{\vec{k}} \approx 0.43$, $A_{\text{ph}}\approx 0.35$, $v_{g} \approx 0.43 c/\sqrt{\epsilon_b}$.
Near the excitonic to photonic crossover $\hbar\omega_{\vec{k}} = \hbar\omega_e$, $\Delta E_{\vec{k}} \approx 0.45$ eV but the mode is not longer confined.
Recall that the emission energy of a localized exciton is peaked at $\hbar\omega_e - \hbar\omega_s$ when the local phonons are taken into account. These quantitative results confirm that the spontaneous emission indeed falls largely into the trapped modes with $\Delta E_{\vec{k}} \approx 2\hbar\omega_s \approx 0.24$ eV, relative to $\hbar\omega_e + \hbar\omega_s$ as explained earlier.

Going back to the main hypothesis. we next investigate the potential decay avenues once a polariton is trapped.  An obvious path is the recapture of the polariton by another donor molecule. The process involves two competing factors, the large number of molecules $N$ and a high thermal activation barrier $\sim 2\hbar\omega_s$ to overcome.  An estimated from Eq.~\cmmnt{(\ref{S_reabsorption})}(S27) in the supplementary information\cite{supplementary} gives the reabsorption rate $\Gamma_r \approx 9\times 10^{-5}\omega_e$ at room temperature, a result to be  revisited below.

Local phonons also induce stochastic fluctuations on physical properties such as the local excited energy. Since a coherent SW state is a superposition of the excited states across the entire crystal, their root mean square averages will be in the order of $1/\sqrt{N}$ ($\sim 2 \times 10^{-5}$). Thus their effects to the spectral broadening of the coherent polariton states are negligible.

The interactions with non-local lattice (acoustic) phonons can be significant. The exciton composition in a polariton can scatter with the latter via long-range dipolar fields. We can nevertheless estimate the effects from the attenuation data measured on a waveguide made of the same materials. The optical loss in Ref.~\cite{ADFM:ADFM201700332} was found to be $0.033$ dB$/\mu$m on a $5 \times 5 \times 50$ $\mu$m crystalline rod, fabricated with the same material via a similar method. Taking an average group velocity $\sim c/(3\sqrt{\epsilon_b})\approx c/4$, cf. Eq.~\cmmnt{(\ref{S_groupv})}(S20), the light ($567$ nm) completes about 7 periods per $\mu$m, which gives a quality factor $Q\approx 0.6\times 10^4$ or a decay time $\tau\sim 3$ ps. Since the value is an all-in-one result and is about the same as one would have obtained from the reabsorption alone, the damping by phonon scattering is likely not important in the current context.

For a dielectric cavity with dimensions close to the underlying photon wavelength, the intrinsic factor limiting the resonance quality comes from the runaway of evanescent waves on the cavity surfaces, caused by surface curvatures or cavity edges (e.g. \cite{Wysin:06}). The off-surface decay length scales as $\sim (\vec{k}_{\|}^2-\epsilon_s k_e^2)^{-1/2}$. Both the surface layer and more crucially the group velocity on the wave propagation should be significantly reduced as the photon composition gets smaller, which in turn heavily suppresses the leaking of the evanescent waves. Though a robust estimate remains a challenge, an indirect inference may be conducted here.

Our indirect estimate is based on the observation that the polariton confinement can also prolong the fluorescence lifetime of an exciton in the nanocrystal, relative to that of an excited monomer in solution. This happens when the reabsorption rate $\Gamma_r$ is larger or comparable to the escape rate $\Gamma_p$ on a trapped polariton. Similar situation was reported in Ref.~\cite{1402-4896-50-5-023} where longer lifetimes of excitations generated by passing $\alpha$-particles inside a molecular crystal were observed.
We can establish a relationship between $\Gamma_r$ and $\Gamma_p$, using available experimental data. The relevant analysis is carried out in Part C of the supplementary information\cite{supplementary} and we get $\Gamma_p \approx 0.6 \Gamma_r$. It translates to an escape time $\tau\sim 5$ ps for the captive polariton. The number is in line with the waveguide data above and with the report (a $2\times 10^{-4}$ eV damping factor) in Ref.~\cite{PhysRevLett.105.067401}. The latter serves an excellent reference for it has roughly the same crystal dimensions, energy spectra, and exciton-photon coupling strength.

We can finally move on to the second part of the problem, namely to verify that our proposed mechanism can adequately account for the experimentally observed energy transfer rate. Doping of BF$_{2}$cna or BF$_{2}$dan creates a new escape path for the confined polariton. The transition is again bridged by the small photon composition $A_{\text{ph}} \sim 15\%$ in the captive mode. The transfer rate to acceptors $\Gamma_a$ is calculated in Part D of the supplementary information\cite{supplementary}. A good estimate gives $\Gamma_a = \Gamma_0\times(N_a/N)$ with $N_a$ being the number of acceptors and $\Gamma_0\approx \omega_e$. The linear dependence fits with the experimental observation. Note that the process is often referred to as ``trivial transfer'' in the literature and can be observable at a very low doping of the guest molecules (e.g. $5\times 10^{-6}$mole/mole in Ref.~\cite{ADMA:ADMA201305913}).

Can the energy transfer rate be limited by the length of a nanocrystal when the group velocity is reduced and the escape time is short? Take a reasonable lower bound $v_g\approx c/10$ for an estimate.  A polariton can travel over a length $\sim 5$ $\mu$m in $\sim 100 \times 2\pi/\omega_e\approx 0.2$ ps $\ll 5$ ps. It apparently does have sufficient time to find an acceptor within the nanocrystal. Needless to say the crystal dimensions cannot be too small either as the evanescent waves, which does not interact with the acceptors, can become significant. Note that the transfer rate $\Gamma_a$ competes against the escape rate $\Gamma_p$ of the intermediate polariton out of confinement. Most importantly, 50\% energy transfer from donors to acceptors should occur at $\Gamma_a \approx \Gamma_p$  which corresponds to $N_a/N \approx 0.6 \Gamma_r/\Gamma_0 \approx 5 \times 10^{-5}$. This result, obtained with no adjustable parameters, matches near exactly those reported in Figure 3 of Ref.~\cite{ANIE:ANIE201510503}.

To summarize, both qualitative and quantitative analyses allow us to establish, to an excellent degree of confidence that the self confinement of excitonic polaritons is the likely mechanism behind the ultra high efficiency of energy transfer in the artificial light-harvesting system presented in Ref.~\cite{ANIE:ANIE201510503}. Theoretical calculations based on the mechanism account for most of the observations reported. Other experiments based on organic nanocrystals, such as the one reported in Ref.~\cite{doi:10.1002/anie.201803546}, can possibly be explained via the same mechanism.

Can we test the hypothesis by further experiments? We propose that an infrared absorption measurement may be able to detect the possible transition from the lower to the higher branch of the dispersion. The latter will effectively kick a trapped polariton out of confinement. Hence some abnormal profiles, in both the infrared absorption and the visible emission spectra, ought to show up when the crystals are placed under UV illumination. Finally, our study certainly raises an intriguing question:  Can a similar resonant quantum confinement play a key role in the real photosynthesis?

This work was supported in part by the National Natural Science Foundation of China No. 81473105 (MJX) and No. 16Z103060007 (PA).


%

\end{document}


\title{Supplementary Information for \protect\\ Resonant confinement of excitonic polariton and ultra-efficient light harvest in artificial photosynthesis}
\author{Yong-Cong Chen}
\email{chenyongcong@shu.edu.cn}
\affiliation{Shanghai Center for Quantitative Life Sciences \& Physics Department, Shanghai University, Shanghai 200444, China}
\author{Bo Song}
\affiliation{Shanghai Key Lab of Modern Optical System, School of
Optical-Electrical Computer Engineering, University of Shanghai for Science and Technology, Shanghai 200093, China}
\author{Anthony J. Leggett}
\affiliation{Shanghai Center for Complex Physics, Department of Physics and Astronomy, Shanghai Jiao Tong University, Shanghai 200240, China}
\affiliation{Department of Physics, University of Illinois at Urbana-Champaign, Urbana, IL 61801, USA}
\author{Ping Ao}
\affiliation{Shanghai Center for Quantitative Life Sciences \& Physics Department, Shanghai University, Shanghai 200444, China}
\author{Xiaomei Zhu}
\email[Corresponding author: ]{xiaomeizhu@shu.edu.cn}
\affiliation{Shanghai Center for Quantitative Life Sciences \& Physics Department, Shanghai University, Shanghai 200444, China}

\begin{abstract}
This supplementary information covers additional algebra needed for the main article published in \cite{mainwork}.
It consists of the following parts:
A. The standing-wave solutions of Maxwell equations in a rectangle cavity of homogeneous dielectric medium.
B. Diagonalization of the photon-exciton coupling Hamiltonian in the rectangle cavity.
C. Escape rate of the intermediate polariton out of confinement.
D. Transition rate from a trapped exciton-polariton to doped acceptors.
\end{abstract}
\maketitle


\renewcommand{\vec}[1]{\mathbf{#1}}
\newcommand{\bmu}{\mbox{\boldmath $\mu$}}

\renewcommand{\theequation}{S\arabic{equation}}

\renewcommand{\thesection}{\Alph{section}}

\section{Standing-wave solutions in a rectangle dielectric medium}
Consider the electro-magnetic waves inside a rectangular isotropic dielectric medium $L_1 \sim L_2 \sim \lambda_e$, $L_3 \sim 10\lambda_e$, with $\lambda_e\approx 500$ nm being the (vacuum) wavelength of the exciton emission. The dimensions are chosen to resemble the nanocrystals used in Ref. \cite{ANIE:ANIE201510503}. Let the electric field
\begin{equation}\label{S_maxwell0}
 \vec{E}(\vec{r}, t) = \vec{U}_{\vec{k}\alpha}(\vec{r})\cos(\omega_{\vec{k}} t), \;\;\;  \epsilon_b\omega_{\vec{k}}^{2} = \vec{k}^{2}c^{2}
\end{equation}
be a standing wave inside the cavity with the background relative permittivity $\epsilon_b$ excluding the contribution from the excitons. And
\begin{eqnarray}\label{S_maxwell}
\nonumber  \vec{U}_{\vec{k}\alpha}(\vec{r}) &=& (u_{x\vec{k}\alpha}U_{x\vec{k}}, u_{y\vec{k}\alpha}U_{y\vec{k}}, u_{z\vec{k}\alpha}U_{z\vec{k}})\\
\vec{u}_{\vec{k}\alpha}&=& (u_{x\vec{k}\alpha}, u_{y\vec{k}\alpha}, u_{z\vec{k}\alpha}) \;\;\; \alpha = 1, 2
\end{eqnarray}
where $\vec{u}_{\vec{k}\alpha}$'s are the two polarization vectors and the wave vector $\vec{k}= (k_x, k_y, k_z)$ is limited to the first octant
\begin{align}\label{S_polarization}
\nonumber \vec{k} =(\pi l_1/L_1, \pi l_2/L_2, \pi l_3/L_3,) \;\;\; & l_i = 0, 1, 2, 3, \dots \\
          \tilde{\vec{k}} = \vec{k}/|\vec{k}| = \vec{u}_{\vec{k}1} \times \vec{u}_{\vec{k}2}\;\;\;
        & \vec{u}_{\vec{k}\alpha} \cdot \vec{u}_{\vec{k}\alpha'} = \delta_{\alpha\alpha'}
\end{align}
The wave functions $U_{\gamma\vec{k}}(\vec{r})$ ($\gamma=x, y, z$) themselves take the form
\begin{eqnarray}\label{S_maxwell1}
\nonumber  U_{x\vec{k}}(\vec{r}) &=& \sqrt{8/V}\sin(k_x x + \delta_x)\cos(k_y y + \delta_y)\cos(k_z z + \delta_z) \\
\nonumber  U_{y\vec{k}}(\vec{r}) &=& \sqrt{8/V}\cos(k_x x + \delta_x)\sin(k_y y + \delta_y)\cos(k_z z + \delta_z) \\
           U_{z\vec{k}}(\vec{r}) &=& \sqrt{8/V}\cos(k_x x + \delta_x)\cos(k_y y + \delta_y)\sin(k_z z + \delta_z)
\end{eqnarray}
where $\delta_{\gamma}$'s are constant phase shifts and the normalization factor $\sqrt{8/V}$ can be different when some $k_{\gamma}$'s are zero.
Clearly Eq.~(\ref{S_maxwell1}) satisfies the transverse wave constraint $\nabla\cdot\vec{E} = 0$ as expected. Special case $\delta_{\gamma} = 0$ gives the solutions for $\vec{E}_{\perp} = 0$  boundary condition on all the surfaces, whereas $\delta_{\gamma} = \pi/2$ corresponds to the $\vec{E}_{\|} = 0$ solutions. Both can be found in Chap 13 of Ref.~\cite{Morse_Feshbach}.

The standing waves (\ref{S_maxwell}) are linear superpositions of 8 plane waves with mirror reflections of $\vec{k}$ in some particular way.  A wave is trapped inside when the tangent component of the wave vector
\begin{equation}\label{S_criteria}
\vec{k}_{\|}^2 c^2 > \epsilon_s (E_{\vec{k}}/\hbar)^2 
\end{equation}
holds on all the surfaces. In the above $\epsilon_s$ is the dielectric constant of the surrounding medium and $E_{\vec{k}}$ is the lower-branch energy of the exciton-polariton when the photon-exciton coupling is taken into account, cf. Eq.~(\ref{S_isotropic}). More explicitly, we have
\begin{eqnarray}\label{S_criteria1}
  \nonumber (k_x^2+k_y^2), (k_y^2+k_z^2), (k_z^2+k_x^2) & \ge & \epsilon_s (E_{\vec{k}}/\hbar)^2 \approx \epsilon_s k_e^2 \\
  k_e = \omega_e/c &=& (2\pi)/{\lambda_e}
\end{eqnarray}
since $E_{\vec{k}}$ is capped by the exciton energy $\hbar\omega_e$ in question.

As pointed out in the main article\cite{mainwork}, we focus in this work on the bound solutions as a result of total internal reflection, with zero field boundary condition at infinity. Now Maxwell's equations require the continuity of the tangent electric field $\vec{E}_{\|}$ and the normal electric displacement field $\vec{D}_{\perp}$ across a medium interface. If the dielectric cavity were an isotropic medium (with large $\epsilon$), the $\vec{E}_{\perp} = 0$ set of solutions would be a good approximation and the amplitude of the evanescent waves outside the cavity would be heavily suppressed. The circumstance here is different. $\vec{D}_{\perp}$ in the current context is always present because the induced exciton polarization aligns to the fixed-oriented dipole moment $\vec{d}$ (see below). On the other hand, when $\vec{E}$ is perpendicular to $\vec{d}$ excitons will not be polarized. Hence for a trapped wave $\vec{D}_{\perp}$ can likely be reduced via suitable adjustment on the phase shifts $\delta_{\gamma}$ in such a way that the average $(\vec{E}\cdot\vec{d})^2$ over the surfaces is minimized.
For instance, if $\vec{d}$ happens to be in the $x$ direction then $\delta_x = 0$ and $\delta_y, \delta_z = \pi/2$ can be the ideal adjustment.
In principle, a bound state can survive when a large portion of the EM energy is kept inside to maintain the strength of the photon-exciton coupling. And when it does, the wave forms (\ref{S_maxwell1}) can reasonably represent the eigen state of the electric field inside.
Ultimately the spatial details of $U_{\gamma\vec{k}}(\vec{r})$ are not crucial as they do not enter the eventual coupling Hamiltonian (\ref{S_interaction1}) over the bulk of the cavity.

\section{Diagonalization of the photon-exciton Hamiltonian}

For quantum mechanical modeling of electro-magnetic interaction with the excitons, one can quantize the macroscopic Maxwell equations using these eigenmodes. On a given $\vec{k}$, the photon Hamiltonian and the electric field operator inside the cavity are then given by (see e.g. \cite{PhysRevA.50.1830}),
\begin{eqnarray}\label{S_quantization}
  \hat{H}_{p\vec{k}} & = & \sum_{\alpha}\hbar\omega_{\vec{k}}(\hat{a}_{\vec{k}\alpha}^{\dag}\hat{a}_{\vec{k}\alpha} + \frac{1}{2})\\
  \hat{\vec{\mathcal{E}}}_{\vec{k}}(\vec{r}) & = & \sum_{\alpha}f_{\vec{k}}(\hat{a}_{\vec{k}\alpha} + \hat{a}_{\vec{k}\alpha}^{\dag})\vec{U}_{\vec{k}\alpha}(\vec{r})
\end{eqnarray}
where $f_{\vec{k}} = \sqrt{\hbar\omega_{\vec{k}}/(2\epsilon_0\epsilon_b)}$ and the phase of $\hat{a}_{\vec{k}\alpha}$ is chosen for symmetric $\hat{\vec{\mathcal{E}}}$.

We can approximate an excited molecule by a harmonic oscillator $\hat{b},\hat{b}^{\dag}$ restricted to its zeroth and first level, $E_1 = \hbar\omega_e\approx 2.5$ eV, namely limiting to the one excitation manifold.
Let the position of the $j$th molecule be $\vec{R}_j$ and the dipole matrix element between the ground and the excited states be $\vec{d}=\langle g|\hat{\vec{r}}|e\rangle$, then we have the exciton-photon interaction Hamiltonian
\begin{eqnarray}\label{S_interaction}
  \hat{H}_{I\vec{k}} & = & \sum_{j, \alpha}ef_{\vec{k}}\left[
  \vec{U}_{\vec{k}\alpha}(\vec{R}_j)\cdot(\vec{d}^{\ast}\hat{b}_{j}^{\dag}\hat{a}_{\vec{k}\alpha} + \vec{d}\hat{a}_{\vec{k}\alpha}^{\dag}\hat{b}_{j})
  \right] \\
\nonumber \empty & \empty & [\hat{b}_i,\hat{b}_j^{\dag}] = \delta_{ij}, \;\;\; i,j = 1, 2, \cdots, N
\end{eqnarray}
A rotating wave approximation (e.g. \cite{HEPP1973360,PhysRevLett.93.226403}) has been used as we focus on the resonance scenario. By taking all molecules with the same matrix element, we have ignored the minor orientation difference between the two groups of molecules in the crystal.

Similar to Refs.~\cite{PhysRev.93.99,PhysRev.112.1555}, we can construct a standing-wave (SW) superposition of the identical electronic states from all the molecules,
\begin{align}\label{S_coherent}
  \hat{b}_{z\vec{k}}^{\dag} &  = \sqrt\frac{V}{N}\sum_j\left[U_{z\vec{k}}(\vec{R}_j)\hat{b}_{j}^{\dag}\right] \\
  \empty & |\phi_{z\vec{k}}\rangle  = \hat{b}_{z\vec{k}}^{\dag}|0\rangle \;\;\; [\hat{b}_{z\vec{k}},\hat{b}_{z\vec{k}}^{\dag}] = 1
\end{align}
Two more SW states $|\phi_{x\vec{k}}\rangle$ and $|\phi_{y\vec{k}}\rangle$ can be likewise defined respectively for the $x$ and $y$ components. The three SW states are mutually orthogonal (for large $N$). Though the SW states from different $\vec{k}$'s can overlap due to over-completeness of the construction.

In terms of the coherent SW states\cite{PhysRev.93.99, HEPP1973360}, the non-interactive (zeroth order) Hamiltonian for a given $\vec{k}$ is simply
\begin{equation}\label{S_hamiltonian}
 \hat{H}_{0\vec{k}} = \sum_{\gamma}\hbar\omega_e\hat{b}_{\gamma\vec{k}}^{\dag}\hat{b}_{\gamma\vec{k}} +
 \sum_{\alpha}\hbar\omega_{\vec{k}}\hat{a}_{\vec{k}\alpha}^{\dag}\hat{a}_{\vec{k}\alpha}
\end{equation}
where $\gamma=x, y, z$ and the interaction term (\ref{S_interaction}) becomes
\begin{equation}\label{S_interaction1}
 \hat{H}_{I\vec{k}} = \sum_{\alpha, \gamma}\left[g_{\vec{k}}
            \tilde{\vec{d}}_{\gamma}^{\ast} \cdot \vec{u}_{\vec{k}\alpha}\hat{b}_{\gamma\vec{k}}^{\dag}\hat{a}_{\vec{k}\alpha} + h.c.\right] \\
\end{equation}
\begin{equation}\label{S_coupling}
 g_{\vec{k}} = e|\vec{d}|f_{\vec{k}}\sqrt{N/V}, \;\;\; \tilde{\vec{d}}_{\gamma} = \vec{d}_{\gamma}/|\vec{d}|
\end{equation}
The scale of the exciton-photon interaction can be determined from the experimental data as is done in the main article\cite{mainwork}. At the resonance energy $\hbar\omega_{\vec{k}}=\hbar\omega_e$, $\epsilon_b\sim 2$, the coupling constant $g_{\vec{k}}$ evaluates to $\sim0.8$ eV.

We now proceed to diagonalize $(\hat{H}_{0\vec{k}}+\hat{H}_{I\vec{k}})$. There is always a decoupled solution $E^L_{\vec{k}} = \hbar\omega_{e}$ that corresponds to the longitudinal wave mode in Ref.~\cite{PhysRev.93.99} or \cite{Morse_Feshbach}. Introduce $E_{\vec{k}} = g_{\vec{k}}\Lambda$ and further dimensionless variables
\begin{equation}\label{S_dimensionless}
\tilde{\Lambda}^2 = (\Lambda-\tilde{\omega}_{e})(\Lambda-\tilde{\omega}_{\vec{k}}),\; \tilde{\omega}_{e} = \hbar\omega_{e}/g_{\vec{k}},\;\tilde{\omega}_{\vec{k}} = \hbar\omega_{\vec{k}}/g_{\vec{k}}
\end{equation}
The equation for the remaining four roots reads after some manipulations,
\begin{equation}\label{S_eigenvalue}
\tilde{\Lambda}^{4}- \sum_{\gamma}|\tilde{\vec{d}}_{\gamma} \times \tilde{\vec{k}}|^2\tilde{\Lambda}^2 + \frac{1}{2}\sum_{\gamma,\gamma'}|(\tilde{\vec{d}}_{\gamma}\times\tilde{\vec{d}}_{\gamma'})\cdot\tilde{\vec{k}}|^2 = 0
\end{equation}
In the above the two polarization vectors have been eliminated via the identities,
\begin{equation*}
\sum_{\alpha}|\tilde{\vec{d}}_{\gamma}\cdot\vec{u}_{\vec{k}\alpha}|^2 \equiv |\tilde{\vec{d}}_{\gamma}|^2-|\tilde{\vec{d}}_{\gamma}\cdot\tilde{\vec{k}}|^2 \equiv |\tilde{\vec{d}}_{\gamma} \times \tilde{\vec{k}}|^2
\end{equation*}
In addition, the compositions of the diagonalized eigenstate
\begin{equation}\label{S_eigenstate}
|E_{\vec{k}}\rangle =
\left[\sum_{\gamma}{\beta}_{\gamma\vec{k}}\hat{b}_{\gamma\vec{k}}^{\dag} +
 \sum_{\alpha}{\alpha}_{\vec{k}\alpha}\hat{a}_{\vec{k}\alpha}^{\dag}\right]|0\rangle
\end{equation}
satisfy
\begin{equation}\label{S_particle}
\sum_{\gamma}\frac{|\beta_{\gamma\vec{k}}|^2}{|\tilde{\vec{d}}_{\gamma}|^2} = \sum_{\alpha}\frac{|\alpha_{\vec{k}\alpha}|^2}{(\Lambda-\tilde{\omega}_{e})^2} = \frac{A_{\text{ph}}}{(\Lambda-\tilde{\omega}_{e})^2}
\end{equation}
where we have introduced a photon composition factor $A_{\text{ph}}$ for the polariton.
Since $\Lambda$ must be real,  $\tilde{\Lambda}^2$ has at least one positive root. Hence there is always a solution to Eq.~(\ref{S_eigenvalue}) such that $\Lambda < \min(\tilde{\omega}_e,\tilde{\omega}_\vec{k})$. 
Take a simple isotropic scenario where all $|\tilde{\vec{d}}_{\gamma}|=1/\sqrt{3}$ then $\tilde{\Lambda}^2 = 1/3$. Define the energy drop $\Delta E_{\vec{k}} = \hbar\omega_e - E_{\vec{k}}$ and the photon-exciton frequency detuning $\Omega_{\vec{k}} = \omega_{\vec{k}}-\omega_e$, the lower branch dispersion reads
\begin{equation}\label{S_isotropic}
\Delta E_{\vec{k}} = \sqrt{\frac{(\hbar\Omega_{\vec{k}})^2}{4} + \frac{g_{\vec{k}}^2}{3}} - \frac{\hbar\Omega_{\vec{k}}}{2}
\end{equation}
Since $\Delta E_{\vec{k}}$ is capped by $\hbar\omega_e$, the magnitude of the related group velocity for a trapped polariton $v_{g}(\vec{k})$ is greatly reduced for large $\Omega_{\vec{k}}$,
\begin{equation}\label{S_groupv}
v_g(\vec{k}) = \frac{|\nabla_{\vec{k}}E_{\vec{k}}|}{\hbar} = \frac{c}{\sqrt{\epsilon_b}}\left[\frac{\Delta E_{\vec{k}} + G}{2\Delta E_{\vec{k}} + \hbar\Omega_{\vec{k}}}\right]
\end{equation}
where $G = g_{\vec{k}}^2/(3\hbar\omega_{\vec{k}}) \approx 0.08$ eV, independent of $\vec{k}$. Numerically the bracket on the right-hand side of Eq.~(\ref{S_groupv}) is, as shown in the main article\cite{mainwork}, roughly the same as $A_{ph}$ from Eq.~(\ref{S_particle}).

\section{Escape rate of the intermediate polariton}
As discussed in the main article\cite{mainwork}, an exciton in the current context first decays into an intermediate exciton-polariton trapped inside by total internal reflection. The intermediate polariton can possibly be reabsorbed by another donor molecule in the crystal or escape into the open space. If the reabsorption rate $\Gamma_r$ is larger or comparable to the escape rate $\Gamma_p$,  the confinement can significantly prolong the fluorescence lifetime of the excitons. We can use this property to establish a relationship between $\Gamma_r$ and $\Gamma_p$.

For simplicity we will take the reabsorption process as the dominant cause for the observed delay in the fluorescence decay process. This can in fact be roughly justified based on the experimental data.
From Table 1 of Ref.~\cite{ADFM:ADFM201700332}, an exciton on a monomer in DCM solution has the fluorescence lifetime $\tau_f = 2.12$ ns, the quantum yield $\Phi_f=0.80$. In the nanocrystal, from Table S1 of Ref.~\cite{ANIE:ANIE201510503}, the average lifetime $\tau_f' = 4.35$ becomes longer and the quantum yield $\Phi_f'=0.37$  (also reported in Ref.~\cite{ADFM:ADFM201700332}) is lower. We see no significant difference between the non-radiative decay rates $(1-\Phi_f)/\tau_f$ in the solution and $(1-\Phi_f')/\tau_f'$ in the crystal. Hence one can simply assume that the non-radiative decay process remain unchanged in the two systems.

Both the emission from a localized exciton into a ballistic exciton-polariton or the reabsorption of the latter by another donor molecule are likely engaged with local phonons and their thermal activities. It would be difficult to combine the two processes into a unified quantum mechanical calculation. Instead, we can resort to a simple population dynamics that describes the statistical outcome on a large number of nanocrystals and measurements. Let $A(t)$, $B(t)$ be the respective populations of localized excitons and extended exciton-polaritons then
\begin{eqnarray}\label{S_two-step}
\nonumber \dot{A}(t) & = & -(1/\tau_f)A(t) + \Gamma_r B(t) \\
\dot{B}(t) & = & (\Phi_f/\tau_f)A(t)  - (\Gamma_r + \Gamma_p) B(t)
\end{eqnarray}
Note that the term $(1/\tau_f)A(t)$ includes the non-radiative decay mentioned above.

The solutions of Eq.~(\ref{S_two-step}) take the form $\exp(-\omega t)$ and we can solve the eigen values for $\omega$. The inverse of the root closer to $1/\tau_f$ will correspond to the prolonged lifetime $\tau_f'$. For $\Gamma_p, \Gamma_r \gg 1/\tau_f$, the eigen equation of the two-step decay (\ref{S_two-step}) gives
\begin{equation}\label{S_lifetime}
  (1/\tau_f - 1/\tau_f')(\Gamma_r + \Gamma_p) \cong (\Phi_f/\tau_f)\Gamma_r
\end{equation}
Using the experimental values mentioned above for $\tau_f$, $\Phi_f$ and $\tau_f'$ we get $\Gamma_p \approx 0.6 \Gamma_r$. It translates to an escape time $\tau\sim 5$ ps for the captive polariton with $\Gamma_r$ taken from Eq.~(\ref{S_reabsorption}).
\section{Transition rate to doped acceptors}
Doping of acceptors BF$_{2}$cna or BF$_{2}$dan 
creates a new escape path for the confined polariton. The transition is bridged by the small photon composition $A_{\text{ph}} \sim 15\%$ in the captive mode.
The transition matrix element from a photon state $\hat{a}_{\vec{k}\alpha}^{\dag}|0\rangle$ to an acceptor at $\vec{R}$ (with dipole matrix element $\vec{d}_a$) is
\begin{equation}\label{S_matrix0}
  t_{\vec{k}}(\omega_{\vec{k}}, \vec{R}, \alpha) =   ef_{\vec{k}}\vec{d}_a\cdot \vec{U}_{\vec{k}\alpha}(\vec{R})
\end{equation}
Consider $N_a$ acceptors homogeneously distributed over the crystal. The total transition matrix element $T_{\vec{k}}$ to these acceptors is,
\begin{equation}\label{S_acceptor}
  |T_{\vec{k}\alpha}|^2 = \sum_{j=1}^{N_a} |t_{\vec{k}}(\omega_{\vec{k}}, \vec{R}_j, \alpha)|^2
\end{equation}
The ensemble average over $\vec{R}_j$ and the polarizations gives
\begin{equation}\label{S_matrix1}
\left<|T_{\vec{k}}|^2\right> 
=\frac{e^2f_{\vec{k}}^2N_a}{2V}\sum_{\gamma}|\vec{d}_{a\gamma} \times \tilde{\vec{k}}|^2
\end{equation}
We can further take $\vec{d}_a\approx \vec{d}$ and the bandwidth $\Delta\omega$,  $\omega_{e}/\Delta\omega\approx 10$ for the acceptors, the Fermi golden rule gives the transition rate to the acceptors
\begin{equation}\label{S_transition}
\Gamma_a \approx \frac{2\pi A_{\text{ph}} }{3} \left(\frac{g_{\vec{k}}}{\hbar\omega_e}\right)^2\left(\frac{\omega_e}{\Delta\omega}\right) \omega_e \times \frac{N_a}{N} \equiv \Gamma_0 \times \frac{N_a}{N}
\end{equation}
It turns out $\Gamma_0 \approx \omega_e$.
A similar derivation can be applied to the reabsorption rate $\Gamma_r$ by a donor molecule. It involves
two major factors, the large number of localized states $N$ and a high thermal activation barrier to overcome, hence
\begin{equation}\label{S_reabsorption}
\Gamma_r \approx \Gamma_0 \times \exp(-2\hbar\omega_s/kT)\approx \Gamma_0\times9\times 10^{-5}.
\end{equation}

%